\documentstyle[preprint,eqsecnum,aps]{revtex}
\begin{document}
\draft
\preprint{SOGANG-HEP 236/98, SNUTP 98-057, hep-th/9806112}
\title{Pointlike structure for super p-branes}
\author{Hyun Seok Yang, Inbo Kim and Bum-Hoon Lee}
\address{Department of Physics, Sogang University, Seoul 121-742, Korea}
\maketitle

\begin{abstract}
We present an efficient method to understand the p-brane dynamics in a
unified framework. For this purpose, we reformulate 
the action for super p-branes in the form appropriate to incorporate 
the pointlike (parton) structure of higher
dimensional p-branes and intend to interpret the p-brane dynamics 
as the collective dynamics of superparticles. In order
to examine such a parton picture of super p-branes, 
we consider various superparticle configurations 
that can be reduced from super p-branes,
especially, a supermembrane, and study the partonic structure of
classical p-brane solutions.
\end{abstract}
\baselineskip=20pt
\def\be{\begin{equation}}
\def\ee{\end{equation}}
\def\bea{\begin{eqnarray}}
\def\eea{\end{eqnarray}}
\def\ba{\begin{array}}
\def\ea{\end{array}}
\def\l{\label}
\def\r{\ref}
\def\c{\cite}
\def\n{\nonumber}
\narrowtext

\section{Introduction}

The eleventh dimensions is the highest dimension in which 
supergravity theory can exist, with fields carrying spin $J \le 2$ \c{nahm}. 
In fact, it is the only sensible supersymmetric theory in $d=11$ \c{cjs}. 
It has a membrane as a fundamental
degree of freedom as well as gravitons \c{bstow,aetw}, which 
may come from the massless excitations of a supermembrane. 
Recently, it was shown \c{HT95,Witt95} that
it is the low energy limit of the eleven dimensional M theory. 
The M theory is defined as the strong coupling limit of the type IIA
string theory \c{HT95,Witt95} and 
the double dimensional reduction of an eleven
dimensional supermembrane action yields the Green-Schwarz action of
the type IIA superstring \c{duff1}. 
These lead one to wonder whether a quantum
supermembrane provides an intrinsic definition of M-theory. Moreover,
it was shown \c{bps} that the massless spectrum of a supermembrane in $d=11$
occurring in the sector of a completely collapsed membrane,
i.e., superparticle, corresponds to the supergravity multiplet. 

But the principal objection to this reasoning is that the spectrum is
continuous \c{whn,dln}, 
which would preclude a particle interpretation. It is known that, 
unlike string theory, membrane theory encounters new
divergences coming from an infinite number of internal degrees of
freedom. In order to make the supermembrane dynamics well-defined, we
need to have some kind of regularization in a supersymmetric way. 
Such a regularized description, so-called matrix theory, is given 
in a light-cone gauge by $U(N)$ supersymmetric Yang-Mills 
quantum mechanics \c{whn} and its underlying
spacetime geometry is noncommutative at short distances. The classical
spacetime geometry is a sensible concept only in a long distance
regime. Thus, the spectrums of short distance physics may be
dramatically changed \c{town} due to noncommutativity of spacetime.  

The parton model of hadrons in the late 1960s was originally
developed to describe the properties of high energy collisions of
hadrons and later incorporated into the fabric of
the quantum chromodynamics, generally accepted relativistic parton
model of hadrons. In an infinite momentum frame in which 
partons are in extreme relativistic motion, 
the internal motions of the partons and the rate at
which they interact with each other are slowed down (frozen) because of
the relativistic time dilatation effect and the Fock space vacuum
becomes extremely simple with the nonrelativistic 
nature of underlying dynamics \c{wks}. 
The matrix regularization of supermembrane dynamics 
attempts to develop the theory based on the idea that 
the supermembrane is made of smaller entities, partons \c{whn}.
In addition, the Matrix theory for a complete nonperturbative
formulation of M theory explicitly incorporates the parton picture in
terms of D0-branes in infinite momentum frame \c{BFSS,Matrix}. 
When this is done, the spatial coordinates of the N D0-branes 
are represented by $N \times N$ Hermitian matrices.

The recent picture of M theory tells us that 
strings, membranes and other extended p-branes hold an equal rank 
as nonperturbative spectrums \c{Matrix,schtown}. 
Recently, the ordinary string theory as a
first quantized description was reformulated as the Matrix string theory, 
the Matrix theory compactified on a tiny circle, where it was shown 
that it provides a description of the Hilbert space 
of {\it second} quantized string theory \c{string}. 
In analogy with the
quark picture that appeared to unify many ``fundamental'' hadrons, 
it may be reasonable to consider p-branes as the composites of smaller
entities. It is thus desirable to reformulate in a unified framework 
the p-brane dynamics as the dynamics for possible constituents 
as the Matrix model for M theory \c{BFSS}.

In this paper, we construct the Barbour-Bertotti-Schild action 
\c{bb,se} for super p-branes in order to incorporate 
the point-like (parton) structure of higher
dimensional p-branes and intend to interpret the p-brane dynamics 
as the collective dynamics of superparticles. In order
to examine the parton picture of super p-branes, 
we consider various superparticle configurations 
that can be reduced from super p-branes, especially, supermembrane 
and study the partonic structure of classical
p-brane solutions. Finally, we give some comments on the
matrix formulation of the supermembrane from 
the viewpoint of composites of point-like entities.  

\section{Super p-brane Action}

For the purpose of illustrating the point-like structure of p-branes,
in this section, we will first consider the Green-Schwarz action
\c{GS} of a d-dimensional supermembrane and then super p-brane. 
The action for the supermembrane in flat superspace is \c{bstow,aetw}
\be
\l{maction}
I=-T_M\int d^3 \xi \left\{\sqrt{-g(X,\theta)}+i\varepsilon^{ijk}
\left(
\frac{1}{2}\partial_i X^\mu(\partial_j X^\nu -i{\bar
\theta}\Gamma^\nu\partial_j \theta)-\frac{1}{6}{\bar \theta}
\Gamma^\mu\partial_i \theta  {\bar \theta}\Gamma^\nu\partial_j \theta
\right)
{\bar \theta}\Gamma_{\mu\nu}\partial_k \theta \right\},
\ee
where $X^\mu(\xi)$ and $\theta(\xi)$ denote the superspace coordinates
of the membrane parameterized in terms of world volume parameters 
$\xi^i\;(i=0,1,2)$. Here $T_M$ is a membrane tension proportional 
to $l_P^{-3}$ and we will take the unit $T_M=1$. 
The metric $g_{ij}(X,\theta)$ is the induced metric
on the world volume
\be
\l{metric}
g_{ij}=E^\mu_iE^\nu_j \eta_{\mu\nu},
\ee
where $E_i^\mu$ are certain supervielbein components tangential to the
worldvolume defined by
\be
\l{vielbein}
E^\mu_i=\partial_i X^\mu-i{\bar \theta}\Gamma^\mu \partial_i\theta
\ee
and $\eta_{\mu\nu}$ is the flat $d=11$ Minkowski metric.  
The action (\r{maction}) is invarint under spacetime supersymmetric
transformations 
\be
\l{susytf}
\delta X^\mu = i{\bar \epsilon}\Gamma^\mu \theta,\;\;\;
\delta \theta =\epsilon.
\ee
Note that the above invariance is associated with the crucial 
gamma matrix identity 
\be
\l{gammaid}
{\bar \psi}_{[1}\Gamma^\mu \psi_2{\bar \psi}_3
\Gamma_{\mu\nu}\psi_{4]}=0
\ee
only satisfied for $d=4,\,5,\,7$ and $11$ \c{bstow,aetw}.

We would like to rewrite the first term represented as 
the Nambu-Goto type as the
following Schild type action \c{se}:
\be
\l{schild}
-\int d^3 \xi \sqrt{-g(X,\theta)}=
\frac{1}{2}\int d^3 \xi e\left( \frac{1}{3!e^2}(\varepsilon^{ijk}
E_i^\mu E_j^\nu E_k^\rho)^2-1\right).
\ee
Using the equation of motion about the auxiliary field $e$, i.e., 
$e=\sqrt{-(1/3!)(\varepsilon^{ijk}E_i^\mu E_j^\nu E_k^\rho)^2}
=\sqrt{-\mbox{det}g_{ij}}$, it is easy to show that the original 
Nambu-Goto type action can be recovered.

We assume that the topology of the membrane is fixed to be $\Sigma
\times {\bf R}$, with $\Sigma$ a compact two manifold, so that the
three coordinates of the world volume, $\xi^i$, are broken down into
$\xi^0=\tau$ and $\xi^a=\sigma^a,\;a=1,2$.\footnote{The 2+1 splitting
corresponds to a gauge fixing to put shift vectors $N^a$ of 
world volume metric to zero \c{bst}.} 
We introduce a two
dimensional induced metric on $\Sigma$ defined by 
\be
\l{q}
q_{ab}=E_a^\mu E_b^\nu \eta_{\mu\nu},\;\;\;
\varepsilon^{ab}=-\varepsilon^{0ab}.
\ee 
Note that
\be
\l{ee}
\varepsilon^{ab}\varepsilon^{cd}=q(q^{ac}q^{bd}-q^{bc}q^{ad}),
\ee
where $q^{ab}$ is the inverse of $q_{ab}$, i.e., 
$q^{ac}q_{cb}=\delta^a_b$ and $q=\mbox{det}q_{ab}$.

The action (\r{schild}) can then
be rewritten as the Barbour-Bertotti-Schild (BBS) type \c{bb,se} 
appropriate to
incorporating the partonic picture of supermembrane
\be
\l{bbs}
I_{BBS}=
\frac{1}{2}\int d\tau d^2\sigma \sqrt{q} [{\tilde e}^{-1}
({\dot X}^\mu
-i{\bar \theta}\Gamma^\mu {\dot \theta})G_{\mu\nu}
({\dot X}^\nu-i{\bar \theta}\Gamma^\nu {\dot \theta})-{\tilde e}],
\ee  
where dot means the derivative with respect to the time-like
parameter $\tau$ and ${\tilde e}=e/{\sqrt{q}}$. 
Here we have introduced the ``manifold Poisson brackets'' (MPB) 
using the simplectic
structure $\varepsilon^{ab}/\sqrt q$ on the two manifold $\Sigma$:
\be 
\l{mpoisson}
\langle f,g \rangle
=\frac{\varepsilon^{ab}}{\sqrt{q}}\partial_a f \partial_b g
\ee
for $C^\infty(\Sigma)$ functions $f$ and $g$. 
Note that this simplectic structure on $\Sigma$, 
which is the cotangent bundles of configuration space $\Sigma$, 
is uniquely defined by the metric and orientation of $\Sigma$ \c{arnold}. 
Our definition of the MPB manifestly respects the full diffeomorphism
group of $\Sigma,\;Diff(\Sigma)$, and satisfies the Jacobi identity
\be
\l{jacobi}
\langle\langle f, g\rangle, h\rangle+\langle\langle g,h\rangle, f\rangle
+\langle h\langle, f\rangle, g\rangle=0
\ee
for $C^\infty(\Sigma)$ functions $f,\,g$ and $h$.
The metric $G_{\mu\nu}$ on the configuration space
of the embeddings $X^\mu(\sigma^a)$ and $\theta(\sigma^a)$ 
is given by\footnote{The Lorentz indices such as $\mu$ and $\nu$ are
raised and lowered by using the metric $\eta_{\mu\nu}$.}
\be
\l{G}
G_{\mu\nu}=\eta_{\mu\nu}+h_{\mu\nu},
\ee
where $h_{\mu\nu}$ is a useful quantity defined as
\be
\l{h}
h_{\mu\nu}=\langle E_\mu, E^\rho\rangle\langle E_\rho, E_\nu\rangle
=-q^{ab}E_{a\mu} E_{b\nu}
\ee
and used the abbreviated notation
\be
\l{pbee}
\langle E_\mu, E_\nu\rangle =\langle X_\mu,X_\nu\rangle
-i{\bar \theta}\Gamma_\mu\langle \theta, X_\nu\rangle
+i{\bar \theta}\Gamma_\nu \langle\theta, X_\mu \rangle 
+{\bar \theta}\Gamma_\mu
\langle \theta,{\bar \theta}\rangle\Gamma_\nu \theta.
\ee 

The metric $h_{\mu\nu}$ and $G_{\mu\nu}$ {\it induced} by neighboring
superparticles, by the definition of $q_{ab}$, satisfy 
the following identity, respectively,
\bea
\l{1h}
&& h_{\mu\lambda}h^{\lambda\nu}=-{h_\mu}^\nu,\\
\l{1}
&& {G_\mu}^\lambda{G_\lambda}^\nu={G_\mu}^\nu.
\eea  
The Eq. (\r{1}) implies that the metric ${G_\mu}^\nu$, indeed, acts as a
kind of projection operator in the target space ${\bf R}^{d-1,1}$.
In addition, we have the important identity related with $Diff(\Sigma)$ 
constraints generating the
reparameterization of the membrane surface
\be
\l{diff}
E^\mu_a G_{\mu\nu}=0,
\ee
which can be directly derived from the definition (\r{G}) of
$G_{\mu\nu}$. From the Eq. (\r{diff}), 
one can obtain the relation $q_{ab}=-E_a^\mu h_{\mu\nu}
E_b^\nu$, which is consistent with the Eq.(\r{h}).

The action (\r{bbs}) is also invariant under the local
reparameterization, $\tau \rightarrow f(\tau)$, provided that the
auxiliary field ${\tilde e}$ (a sort of ``metric'' along the particle
worldline) transforms as ${\tilde e} \rightarrow {\tilde e}
(df/d\tau)^{-1}$. This reflects that there is no intrinsic preferred
time variable on the membrane. 
Note that, from the equation of motion with respect to the auxiliary field
${\tilde e}$,
\be
\l{e}
{\tilde e}=\sqrt{-({\dot X}^\mu -i{\bar \theta}\Gamma^\mu 
{\dot \theta})G_{\mu\nu}({\dot X}^\nu-i{\bar \theta}\Gamma^\nu {\dot \theta})},
\ee 
we can easily recover the usual Barbour-Bertotti form for 
the membrane as well \c{smolin,am}.

Let us interpret the BBS action (\r{bbs}) as follows. 
The BBS action takes the form of superparticles with unit
mass continuously distributed on the two manifold $\Sigma$ 
moving in a background spacetime metric $G_{\mu\nu}$. 
We would like to interpret the supermembrane as the composite of the 
superparticles bound each other by the surface tension and 
influenced by the {\it effective
gravitational potential $G_{\mu\nu}$}, so, in this sense, the
superparticles play a role of (classical) partons of supermembrane. 
Similarly, we
may consider the supermembrane as the configuration of a fluid
evolving from a fixed initial configuration. 
We can then consider the flow of a nonviscous compressible 
fluid on the region $\Sigma$ moving along the timelike geodesic 
defined by the metric $G_{\mu\nu}$. 
Such a fluid is described by a curve 
$\tau \rightarrow g_\tau$, where the diffeomorphism $g_\tau$ is the
map which carries every particle of the fluid from the place it was at
time 0 to the place it is at time $\tau$. 
From this picture, we see that the classical mass $M$ of the static membrane 
is a sum of the mass of constituent superparticles: i.e., 
\be
\l{mm}
M=T_M \int_{\Sigma}d^2\sigma \sqrt q.
\ee

Recall that the Wess-Zumino term in Eq. (\r{maction}) generates a 
femionic gauge symmetry, so-called $\kappa$-symmetry, which allows us 
to match the Bose and Fermi degrees of freedom. This fermionic gauge 
invariance of the supermembrane is only possible for specific number
of spacetime dimensions, i.e., for $d=4,\,5,\,7$ and $11$. 
The Wess-Zumino action which is independent of the world volume metric
is rewritten as \c{bstow,aetw}
\be
\l{WZ}
I_{WZ}=-\frac{1}{6}\int d^3 \xi \varepsilon^{ijk}E^A_iE^B_jE^C_k B_{CBA},
\ee
where $E_i^A=(E^\mu_i, E_i^\alpha)$ with $E_i^\alpha=\partial_i\theta^\alpha$.
The super 3-form $B$ is such that $H=dB$, with all components of $H$
vanishing except
$H_{\mu\nu\alpha\beta}=-2i(\Gamma_{\mu\nu})_{\alpha\beta}$. 
The gamma matrix identity of Eq. (\r{gammaid}) is nothing
but the Bianchi identity $dH=d^2B=0$ from which the brane scan comes in. 
Solving for $B$, one finds
\be
\begin{array}{ll}
\l{B}
 B_{\mu\nu\rho}=0,&
B_{\mu\nu\alpha}=i({\bar \theta}\Gamma_{\mu\nu})_\alpha,\\
B_{\mu\alpha\beta}=-({\bar \theta}\Gamma_{\mu\nu})_{(\alpha}
({\bar \theta}\Gamma^\nu)_{\beta)},&
B_{\alpha\beta\gamma}=i({\bar \theta}\Gamma_{\mu\nu})_{(\alpha}
({\bar \theta}\Gamma^\mu)_{\beta}({\bar \theta}\Gamma^\nu)_{\gamma)}.
\end{array}
\ee

Since the local $\kappa$-symmetry eliminates half of the $\theta$
fermionic modes, it is involved with some kind
of the projection operator $\frac{1}{2}(1\pm \Gamma)$, where the
function $\Gamma$ is defined by
\be
\l{po}
\Gamma=\frac{1}{6e}\varepsilon^{ijk}E^\mu_iE^\nu_jE^\rho_k 
\Gamma_{\mu\nu\rho}=-\frac{1}{2{\tilde e}}
E^\mu_0\langle E^\nu, E^\rho\rangle \Gamma_{\mu\nu\rho}
\ee
and satisfies the relation $\Gamma^2=1$ on shell. 

In terms of the 2+1 splitting, the action (\r{WZ}) takes the form
\footnote{We are now taking an analogy with electrodynamics, where 
the point particle with charge $q$ is interacting with one-form
potential $A$ defined on the worldline $\Gamma$ of the particle, 
i.e., $q\int_\Gamma A=q\int_\Gamma d\tau (dX^\mu/d\tau)A_\mu$, 
and electromagnetic one-form $A$ should satisfy the Bianchi 
identity, $dF=d^2A=0$.}
\be
\l{wz}
I_{WZ}=\frac{1}{2}\int d\tau d^2\sigma \sqrt{q}
E^A_0\Pi_A,
\ee 
where the ``external'' field $\Pi_A$ is defined as follows:
\be
\l{tcurrent}
\Pi_A=\langle E^B, E^C \rangle B_{CBA}.
\ee 
Then the full BBS type action of the supermembrane takes the following
form:  
\be
\l{full}
I=\frac{1}{2}\int d\tau d^2\sigma \sqrt{q} [{\tilde e}^{-1}
({\dot X}^\mu -i{\bar \theta}\Gamma^\mu {\dot \theta})G_{\mu\nu}
({\dot X}^\nu-i{\bar \theta}\Gamma^\nu {\dot \theta})-{\tilde e}+E^A_0\Pi_A].
\ee 
Now the above supermembrane action can be interpreted as the 
collective dynamics of superparticles 
or the nonviscous charged fluid composed of the
superparticles (which are charged with respect to the fermionic 
$\kappa$-transformation) under the influence of 
the ``gravitational'' field $G_{\mu\nu}$ and the ``external'' field $\Pi_A$. 
We have seen so far that the fields $G_{\mu\nu}$ and $\Pi_A$ which 
couple to the superparticles are not arbitrary, but highly
constrained by $Diff(\Sigma)$ symmetry, supersymmetry and $\kappa$-symmetry. 

Without doing any gauge fixing, we proceed directly to define the canonical
momenta of the variables $(X^\mu,\theta^\alpha)$: 
\be
\begin{array}{l}
\l{canmom}
P_\mu=\delta I/ \delta {\dot X}^\mu= \sqrt{q}\{{\tilde e}^{-1}
G_{\mu\nu}({\dot X}^\nu -i{\bar \theta}\Gamma^\nu {\dot \theta})+
\frac{1}{2}\Pi_\mu \},\\
P_\alpha=-\delta I/ \delta {\dot \theta}^\alpha= 
-i\sqrt{q}\{{\tilde e}^{-1}({\dot X}^\mu -i{\bar \theta}\Gamma^\mu 
{\dot \theta})G_{\mu\nu}({\bar \theta}\Gamma^\nu)_\alpha+
\frac{1}{2}\Pi^\mu ({\bar \theta}\Gamma_\mu)_\alpha
-\frac{i}{2}\Pi_\alpha \} \\
\;\;\;\;\;=-iP^\mu({\bar \theta}\Gamma_\mu)_\alpha
-\frac{\sqrt{q}}{2}\Pi_\alpha.
\end{array}
\ee
The phase space Poisson brackets of these canonical variables are
the following: 
\be
\begin{array}{l}
\l{pspb}
\{X^\mu(\sigma),P_\nu(\sigma')\}_- =\delta^\mu_\nu
\delta^2(\sigma-\sigma'),\\
\{\theta^\alpha(\sigma),P_\beta(\sigma')\}_+ =\delta^\alpha_\beta
\delta^2(\sigma-\sigma'),
\end{array}
\ee
where the graded Poisson bracket is defined by 
$\{A,B\}_{\pm}= \pm \{B,A\}_{\pm}$
and the brackets are evaluated at equal times. 

Let us collect the canonical constraints imposed on the phase space of
the supermembrane \c{bst}:
\bea
\l{c1}
&&F_\alpha \equiv P_\alpha+iP^\mu({\bar \theta}\Gamma_\mu)_\alpha
+\frac{\sqrt{q}}{2}\Pi_\alpha \approx 0,\\
\l{c2}
&& \varphi_{ab}\equiv q_{ab}- \eta_{\mu\nu} E^\mu_aE^\nu_b \approx 0,\\
\l{c3}
&& \varphi_a \equiv E^\mu_a
\left(P_\mu-\frac{\sqrt{q}}{2}\Pi_\mu \right) \approx 0,\\
\l{c4}
&& \varphi \equiv
\frac{1}{2}\left(P_\mu-\frac{\sqrt{q}}{2}\Pi_\mu\right)
G^{\mu\nu}\left(P_\nu -\frac{\sqrt{q}}{2}\Pi_\nu \right)
+\frac{1}{2}q \approx 0,\\
\l{c5}
&& P\equiv \delta I/ \delta {\dot{\tilde e}} \approx 0,\\
\l{c6}
&& P_{ab}\equiv \delta I/ \delta {\dot q}_{ab} \approx 0.
\eea
Note that all these
constraints directly follow from the definition of the phase space
variables. The constraints (\r{c1})
and (\r{c4}) come from the above definition of the conjugate momenta 
$(P_\mu, P_\alpha)$, where the Eq. (\r{e}) is rendered into 
the form of the constraint (\r{c4}). 
The constraint (\r{c2}) is the definition of the
induced metric on the membrane surface $\Sigma$ and (\r{c3}) is the 
$Diff(\Sigma)$ constraint due to the relation (\r{diff}). 
In fact, the constraints (\r{c2}) can be understood as the secondary 
constraints of the second class constraints (\r{c6}). 
On the other hand, the constraint (\r{c5}) is the first class 
generating the reparameterization, ${\tilde e} \rightarrow {\tilde e}
(df/d\tau)^{-1}$. 
Multiplying the constraints (\r{c1})-(\r{c6}) with the Lagrange
multipliers $\Sigma^\alpha,\, \Lambda^{ab},\,\Lambda^a,\,\Lambda,\,
\lambda$ and $\lambda^{ab}$, respectively, and adding them to 
the Hamiltonian, we obtain the total Hamiltonian
\bea
\l{ham} 
H&=&\int d^2\sigma\{(P_\mu {\dot X}^\mu+P_\alpha {\dot \theta}^\alpha-
{\cal L})+\Sigma^\alpha F_\alpha +\Lambda^{ab}\varphi_{ab}+
\Lambda^a \varphi_a +\Lambda \varphi+\lambda P+\lambda^{ab}P_{ab}\}\n\\
&=&\int d^2\sigma\left[\left(\frac{{\tilde e}}{\sqrt {q}}
+\Lambda \right) \varphi
+\Sigma^\alpha F_\alpha +\Lambda^{ab}\varphi_{ab}+
\Lambda^a \varphi_a +\lambda P+\lambda^{ab}P_{ab}\right].
\eea
In Ref.\c{bst}, Bergshoeff {\it et al.} analyzed the constraint
structure of the eleven dimensional supermembrane and covariantly
classified the constraint algebra. 
It was shown in Ref.\c{bst} that Eqs. (\r{c2}), (\r{c6}) and
$1/2(1-\Gamma)(F
+4iP^{ab}E^\mu_a\Gamma_\mu\partial_b\theta)$ 
(which is an orthogonal part on the $\kappa$-transformation) 
are second class constraints. 

It is generally possible that the Green-Schwarz action 
for any p-brane can be rewritten as the BBS action, which takes 
that of particles continuously distributed on a
p-dimensional surface moving in a nontrivial external background. 
The Green-Schwarz action for super p-brane is \c{bstow,aetw}
\be
\l{paction}
I=-T_{p+1}\int d^{p+1} \xi \{\sqrt{-g(X,\theta)}
+\frac{1}{(p+1)!}\varepsilon^{i_1i_2\cdots i_{p+1}}
E^{A_1}_{i_1}E^{A_2}_{i_2}\cdots E^{A_{p+1}}_{i_{p+1}} 
B_{A_{p+1}\cdots A_2A_1}\},
\ee
where the superspace $(p+1)$-form $B$ is the potential 
for a closed $(p+2)$-form $H=dB$. Possible super p-brane theories
exist whenever there is a closed $(p+2)$-form in superspace. 

As the case of supermembrane, we assume that the topology of 
the p-brane is fixed to be $\Sigma
\times {\bf R}$, with $\Sigma$ a compact p-dimensional manifold, 
so that the $(p+1)$ coordinates of the world volume, 
$\xi^i$, are splitted into
$\xi^0=\tau$ and $\xi^a=\sigma^a,\;a=1,\cdots,p$. 
We introduce a p-dimensional induced metric on $\Sigma$ defined by 
\be
\l{qp}
q_{ab}=E_a^\mu E_{b\mu},\;\;\;
\varepsilon^{a_1a_2\cdots a_p}=-\varepsilon^{0a_1a_2\cdots a_p}.
\ee 
Then the following formula can be found
\bea
\l{eep}
\varepsilon^{a_1a_2\cdots a_p}\varepsilon^{b_1b_2\cdots b_p}=q\det
  \left| \matrix{q^{a_1b_1} & q^{a_1b_2} & \cdots & q^{a_1b_p} \cr
                q^{a_2b_1} & q^{a_2b_2} & \cdots & q^{a_2b_p} \cr
                \vdots     & \vdots     & \ddots & \vdots     \cr
                q^{a_pb_1} & q^{a_pb_2} & \cdots & q^{a_pb_p} \cr}
  \right|           
\eea
where $q^{ab}$ is the inverse of $q_{ab}$, i.e., 
$q^{ac}q_{cb}=\delta^a_b$ and $q=\mbox{det}q_{ab}$.

As a result of these formula we have the BBS action for super p-brane
\be
\l{fullp}
I=\frac{1}{2}\int d\tau d^p\sigma \sqrt{q} [{\tilde e}^{-1}
({\dot X}^\mu -i{\bar \theta}\Gamma^\mu {\dot \theta})G_{\mu\nu}
({\dot X}^\nu-i{\bar \theta}\Gamma^\nu {\dot \theta})
-{\tilde e}+E^A_0\Pi_A],
\ee  
where ${\tilde e}=e/{\sqrt{q}}$ and the ``external'' field $\Pi_A$ 
is defined as follows:
\be
\l{efp}
\Pi_A=\frac{2}{p!}\langle E^{A_1},E^{A_2},\cdots, E^{A_p} \rangle 
B_{A_p,\cdots,A_2,A_1,A}.
\ee  
Here we have introduced the ``manifold multiple bracket''\footnote{This
multiple bracket was introduced a long time ago by Nambu \c{nambu} and the
quantization for the generalized Hamiltonian dynamics was
considered. And the basic principles of canonical formalism for the
Nambu dynamics were presented by Takhtajan \c{takh} and applied to the
relativistic p-brane dynamics by Hoppe \c{hopp}.} 
on the manifold $\Sigma$ extending the previous MPB
\be 
\l{mmb}
\langle f_1,f_2,\cdots, f_p \rangle 
=\frac{1}{\sqrt q}\frac{\partial(f_1,f_2,\cdots, f_p)}
{\partial(\sigma_1,\sigma_2,\cdots, \sigma_p)}
\ee
for $C^\infty(\Sigma)$ functions $f_a$. 
The metric $G_{\mu\nu}$ on the configuration space
of the embeddings $X^\mu(\sigma^a)$ and $\theta(\sigma^a)$ is given by
\be
\l{Gp}
G_{\mu\nu}=\eta_{\mu\nu}+h_{\mu\nu},
\ee
where $h_{\mu\nu}$ is defined as
\be
\l{hp}
h_{\mu\nu}=-(-)^{p(p-1)/2}\frac{1}{(p-1)!}
\langle E_\mu, E^{\mu_1},\cdots, E^{\mu_{p-1}}\rangle
\langle E_{\mu_{p-1}},\cdots, E_{\mu_1}, E_\nu\rangle =
-q^{ab}E_{a\mu} E_{b\nu}.
\ee
The similar formula for the metric $h_{\mu\nu}$ and 
$G_{\mu\nu}$ induced by neighboring
superparticles also  hold true for super p-branes
\bea
\ba{llll}
\l{lgp}
&&h_{\mu\lambda}h^{\lambda\nu}=-{h_\mu}^\nu,\\
&&tr h^n=h_{\mu\nu}{h^\nu}_\rho \cdots {h^\sigma}_\lambda h^{\lambda\mu}
=(-)^n \cdot p,\;\;\; \forall n\geq 1,\\
&&{G_\mu}^\lambda{G_\lambda}^\nu={G_\mu}^\nu,\\
&&trG^n=d-p, \;\;\;\;\forall n\geq 1.
\ea
\eea  
In the next section we will show that p-brane solutions always satisfy
these relations.

We have exactly the same kind of identity as the supermembrane 
related with $Diff(\Sigma)$ 
constraints generating the reparameterization of the p-brane surface
\be
\l{diffp}
E^\mu_a G_{\mu\nu}=0.
\ee
From the Eq. (\r{hp}), one can also obtain the relation 
$q_{ab}=-E_a^\mu h_{\mu\nu}E_b^\nu$.
 
Based on their equivalent canonical structure, 
it is apparent that the super p-brane ($p\ge 1$) 
action (\r{fullp}) will exhibit the same Hamiltonian structure 
as the supermembrane action (\r{full}).

\section{Parton Configurations of Super p-branes}

The parton picture in terms of superparticles is quite different
from those of Matrix theory \c{BFSS} and string bits model 
\c{thorn} where partons are 
described by a matrix transforming in the adjoint representation of
some group $G$, mainly $SU(N)$ or $SO(N)$. Nevertheless, the
formulation based on the idea that higher dimensional extended 
p-branes can be made of smaller entities, superparticles, is quite 
useful to understanding the dynamics of p-branes because the dynamics is
conceptually simple and clear. In this section, we will try to
understand the p-branes in viewpoint of composite of superparticles and
study the parton configurations of p-brane solutions.

\subsection{Superstring and superparticle from supermembrane}
First, we consider a double dimensional reduction of 
eleven dimensional supermembrane,
from which the type IIA superstring 
propagating in $d=10$ can be obtained, 
as shown by Duff {\it et al.} \c{duff1}, 
and superparticle in $d=9$ by a further double dimensional reduction. 
In the present viewpoint, these solutions 
can be derived from the particular configurations of superparticles
preserving the supersymmetry.

The type IIA superstring in $d=10$ considered by Duff {\it et al.} \c{duff1} 
is obtained by a compactification of both the
world volume and the spacetime on the same circle, letting the membrane
tension $T_M$ tend to infinity, but the string tension $T_2=2\pi R_1 T_M$
maintain finite. This corresponds to the configuration of 
superparticles whose line mass density along the compactified circle 
tends to infinity, while the mass density along the extended string
remains finite. This situation can be represented by the following ansatz:
\be
\l{string}
X^{10}=\rho,\;\;\;\partial_\rho X^m =\partial_\rho \theta=0,\;\;\;
m=0,1,\cdots,9.
\ee
Then the ``parton metric'' $h_{\mu\nu},\,G_{\mu\nu},\, q_{ab}$ 
and the external field 
$\Pi_A=(\Pi_m, \Pi_\alpha)$ induced by the superparticle 
configuration (\r{string}) can be found as
\bea
\l{10m}
&&h_{\mu\nu}=\pmatrix{ g_{mn} & 0 \cr
                       0 & -1 \cr}, 
\;\;\;\;
G_{\mu\nu}=\pmatrix{ \eta_{mn}+g_{mn} & 0 \cr
                     0 & 0 \cr},
\;\;\;\;
q_{ab}=\pmatrix{ E^m_{\sigma}E_{\sigma m} & 0 \cr
                     0                    & 1 \cr},\\
\l{10pi}
&&\Pi_m=-\frac{2}{\sqrt q}E^\alpha_\sigma B_{10 m \alpha}=
i\frac{2}{\sqrt q}{\bar \theta}\Gamma_{10}\Gamma_{m}
\partial_\sigma\theta,\n\\
&&\Pi_\alpha=\frac{2}{\sqrt q}(E^m_\sigma B_{10 m \alpha}+
E^\beta_\sigma B_{10\beta\alpha})
=i\frac{2}{\sqrt q}({\bar \theta}\Gamma_{10}\Gamma_{m})_\alpha\,
(\partial_\sigma X^m -i{\bar\theta}\Gamma^m\partial_\sigma \theta)\\
&&\;\;\;\;\;\;\;\;\;
-\frac{1}{\sqrt q}[({\bar \theta}\Gamma_{10}\Gamma_{m})_\alpha
\,{\bar \theta}\Gamma^{m}\partial_\sigma\theta-
({\bar \theta}\Gamma_{m})_\alpha
\,{\bar \theta}\Gamma_{10}\Gamma^{m}\partial_\sigma\theta],\n
\eea
where $\eta_{mn}$ is a ten dimensional Minkowski metric and the
petentials $B_{10\alpha m}$ and $B_{10\beta\alpha}$ 
has been determined by Eq.(\r{B}). 
The reduced supervielbein $E_\sigma^A$ and 
the metric $g_{mn}$ are given by
\bea
\l{10e}
&&E_\sigma^A=(E^m_\sigma, E_\sigma^\alpha)=
(\partial_\sigma X^m -i{\bar
\theta}\Gamma^m\partial_\sigma \theta,\,
\partial_\sigma\theta^\alpha),\n\\
&&g_{mn}=-E_{\sigma m}E_{\sigma n}/q, \;\;\;\; 
q=E^l_{\sigma}E_{\sigma l}.
\eea
One can check that the solution (\r{10m}) manifestly satisfies the
identities (\r{lgp}), i.e., $trh^n=(-)^n\cdot 2$ and $trG^n=9$.

In order to recast the Green-Schwarz action for the type IIA
superstring \c{GS}, the 32 components of Majorana spinor $\theta$ can be
splitted into two Majorana-Weyl spinors in terms of the ten 
dimensional chiral matrix $\Gamma_{10}$
\[\theta_{\pm}=\frac{1}{2}({\bf 1}\pm \Gamma_{10})\theta.\] 
Using these results, we can obtain the BBS action for 
the type IIA superstring\footnote{It can be easily showed that the
other $2+1$ splitting from the supermembrane action (\r{maction}), 
$\xi^2=\rho,\,\xi^a=(\tau,\sigma),\, a=0,1,$ directly
gives the Nambu-Goto action of superstring. In this case, the analogue
of the Eq. (\r{full}) is involved with the derivative with respect to
$\rho$ instead of $\tau$. Thus, 
it is sufficient that we consider only terms 
involved with $G_{1010}$ and $\Pi_{10}$.} 
\be
\l{sbbs}
I=\frac{T_2}{2}\int d\tau d\sigma \sqrt{q} [{\tilde e}^{-1}
({\dot X}^m -i{\bar \theta}\Gamma^m {\dot \theta})G_{mn}
({\dot X}^n-i{\bar \theta}\Gamma^n {\dot \theta})-{\tilde
e}+E^A_0\Pi_A], 
\ee  
where $G_{mn}=\eta_{mn}+g_{mn}$ and $\theta=(\theta_+, \theta_-)$.

From the double dimensional reduction (\r{string}), the Hamiltonian
formulation of the superstring can be also derived from the 
Eqs. (\r{canmom})-(\r{ham}) and the constraint structure of the
superstring is the same as that of the supermembrane 
\c{hfs}. 
Note that the string action (\r{sbbs}) can be rewritten as the
superconformally invariant theory through the Polyakov action 
even though the membrane action we started from can not. 

Consider a further double dimensional reduction of the superstring
constructed by the Kaluza-Klein truncation (\r{string}) of the
supermembrane \c{duff2}. 
The string is then wrapping around another circle 
of radius $R_2$. Thus the membrane has a toroidal topology embedded in a 
spacetime ${\bf R}^9\times S^1 \times S^1$. 
Choosing the $S^1 \times S^1$ to be in the $X^{10}$ and $X^9$
directions and letting the string tension $T_2$ tend to infinity, 
but the static membrane mass (\r{mm}), $M=(2\pi R_2)(2\pi R_1) T_M$, 
maintain finite,\footnote{The classical mass of the toroidal solution
considered here is nonvanishing. This mass can be interpreted as 
the winding energy of the membrane wrapping around the toroidal 
surface or that of the 
string wrapping around the $X^9$-circle. For this reason, the mass is
essentially quantized.} 
the classical solution of this configuration 
can be taken as the following form
\be
\l{particle}
X^{10}=\rho,\;\;X^9=\sigma,\;\;\partial_a X^m=\partial_a \theta=0,
\;\; a\in (\sigma, \rho)\;\; \mbox{and}\;\;m=0,1,\cdots,8.
\ee  
This configuration corresponds to a supermembrane that has completely
collapsed to a point. In fact, we find the parton metric

\bea
\l{9p}
&&h_{\mu\nu}=\pmatrix{ g_{mn}=0 & 0  & 0 \cr
                       0 & -1 & 0 \cr
                       0 & 0  & -1 \cr}, 
\;\;\;\;
G_{\mu\nu}=\pmatrix{  \eta_{mn} & 0 & 0 \cr
                     0 & 0 & 0 \cr
                     0 & 0 & 0 \cr},
\;\;\;\;
q_{ab}=\pmatrix{ 1 & 0 \cr
                 0 & 1 \cr},\\
\l{9pi}
&&\Pi_m=\Pi_\alpha=0.
\eea
For these the supermembrane action reduces to that of superparticle
with mass $M$ 
and propagating in $d=9$ with ${\cal N}=2$ supersymmetries
\footnote{A representation of the $\Gamma$-matrices appropriate to the
$11=9+2$ split that we are making is:
\bea
&&\Gamma^m=\gamma^m\otimes \sigma^3,\;\;\;m=0,1,\cdots,8,\n\\
&&\Gamma^{8+a}={\bf 1}_{16}\otimes \sigma^a,\;\;\;a=1,2,\n
\eea
where the 9-dimensional $\gamma$-matrices satisfy 
\[ \{\gamma^m,\gamma^n \}=2\eta^{mn}.\]}
\be
\l{sp}
I=\frac{M}{2}\int d\tau [{\tilde e}^{-1}
({\dot X}^m -i{\bar \theta}^A\gamma^m {\dot \theta}^A)\eta_{mn}
({\dot X}^n-i{\bar \theta}^B\gamma^n {\dot \theta}^B)-
{\tilde e}], 
\ee
where $\theta^A=(\theta_+, \theta_-)$ are two 16 component 
Majorana spinors in $d=9$. For the case of the superparticle, the
``external fields'' $h_{\mu\nu}$ and $\Pi_A$ dissappear in the
action. As a well-known fact, in the case of point particles, 
there is no need for Wess-Zumino term to realize the $\kappa$-symmetry
\c{siegel} as illustrated in Eq. (\r{9pi}). 

Using the above results, the Hamiltonian formulation for the
superparticle \c{siegel,hfsp} can be also derived from the 
Eqs. (\r{canmom})-(\r{ham}) where the nontrivial constraints come from 
Eqs. (\r{c1}), (\r{c4}) and (\r{c5}), the other constraints 
identically (or strongly) vanish.

\subsection{Pulsating spherical membrane}
Consider a periodic pulsating membrane, originally described 
by Collins and Tucker \c{ct}, where a spherical membrane contracts to a point
and expands again with the opposite orientation. This solution was 
recently reconsidered in the Matrix theory context \c{ktr}, where it was 
argued that, upon gravitational collapse, the spherical membrane
has a possibility to form a Schwarzschild black hole and then decay
quantum mechanically via Hawking radiation. As a simple good example of this
formalism, the dynamics of a spherical membrane can be described 
by the $SU(N)$ Yang-Mills quantum mechanics 
in a light-cone gauge \c{whn,dln,ktr}. 

First, we introduce an parameterization of a unit sphere by coorninates
$\sigma^a=(x, \theta)$ with $-1 \le x \le 1$ and $0\le \theta \le
2\pi$. The embedding Cartesian coordinates on the sphere
\be
\l{shere}
x_1=x,\;\;\;x_2=\sqrt{1-x^2} \sin\theta,\;\;\;x_3=\sqrt{1-x^2} \cos\theta
\ee
obey the $SU(2)$ algebra
\be
\l{su2}
\langle x_i, x_j\rangle 
=\varepsilon_{ijk} x_k,\;\;\; i,\,j,\,k =1,2,3.
\ee

The pulsating spherical membrane is described by setting
\bea
&& X_0(\tau, \sigma^a)=t(\tau),\;\;X_i(\tau, \sigma^a)=r(\tau)x_i(\sigma^a),
\;\;X_4=\cdots=X_{d-1}=0,\n\\
&& \theta=0.
\eea
Using the result (\r{su2}), the parton metric can be found as
\bea
\l{s2}
&& G_{ij}=\pmatrix{ x^2 & x\sqrt{1-x^2} \sin\theta & 
     x\sqrt{1-x^2} \cos\theta \cr
     x\sqrt{1-x^2} \sin\theta & (1-x^2) \sin^2\theta  & 
     (1-x^2) \sin\theta \cos\theta \cr
     x\sqrt{1-x^2} \cos\theta & (1-x^2) \sin\theta \cos\theta & 
     (1-x^2) \cos^2\theta \cr},\\
&& q_{ab}=\pmatrix{ \frac{r^2}{1-x^2} & 0 \cr
                 0 &  r^2(1-x^2) \cr},\;\;\;\;\; q=r^4,\n
\eea
where we have explictly presented only the non-flat spatial 
components of $G_{\mu\nu}$. 
The membrane action (\r{full}) can then be reduced to 
the following simple form: 
\bea
\l{rs2}
I&=&\frac{4\pi}{2}\int d\tau\, r^2[{\tilde e}^{-1}(-{\dot t}^2 
+{\dot r}^2)-{\tilde e}]\n\\
&=& -4\pi \int d\tau\, r^2 \sqrt{{\dot t}^2 -{\dot r}^2},
\eea
where we find that $h_{\mu\nu}$ gives no contribution in the action 
(\r{rs2}) due to the relation $x_i^2=1$. 
The equation of motion coming from the variation $\delta t$ is given
by 
\be
\l{eoms2}
\partial_\tau \left(\frac{r^2\, {\dot t}}
{\sqrt{{\dot t}^2 -{\dot r}^2}}\right)=0.
\ee
The action (\r{rs2}) still has the reparameterization symmetry; 
$\tau \rightarrow f(\tau)$. Using this freedom, let us choose a
synchronous gauge
\be
\l{t1}
t=\tau.
\ee
Then the solution takes the form of energy conservation
\be
\l{ec}
{\dot r}^2+\frac{r^4}{r_0^4}=1,
\ee
where $r_0$ is the radial position at $\tau=0$. 
The bosonic partons perform a pulsating motion 
by the attractive $r^4$ potential \c{ct}. 
Note that the potential proportional to $r^4$ comes from the
time-dependent effective
mass of parton, Eq. (\r{mm}), due to the tension of the membrane. 

It is easy to check
that the equation of motion with respect to $r$ is consistent 
with Eq. (\r{ec}). Thus the dynamics of the spherical membrane is fully
determined by Eq. (\r{ec}) which can be solved in terms of elliptic
functions \c{GR}. 

\subsection{Hoppe-Nicolai Solution}

We will consider more general solutions presented by Hoppe and Nicolai
\c{hn} and, in the Matrix theory context, by Hoppe and Rey \c{hopp2}, 
which describes pulsating and rigidly rotating classical 
surfaces (of arbitrary dimension) embedded into Euclidean
spheres. 

We take a natural ansatz corresponding to the simple motions of
pulsation [described by a radial function $r(\tau)$] 
and rotation [described by a time-dependent
real orthogonal matrix $D(\tau)$],
\be
\l{pa}
X_0(\tau, \Omega)= t(\tau),\;\;\; {\bf X}(\tau, \Omega)
=r(\tau)D(\tau) {\bf m}(\Omega),
\ee
where $\Omega=(\sigma^1,\cdots,\sigma^p)$ stands for the world volume 
parameters of a p-dimensional surface and ${\bf m}(\Omega)$ is a unit
vector
\be
\l{m2=1} 
{\bf m}^2(\Omega)=1.
\ee
Then ${\bf X}(\tau, \Omega)$ has the simple interpretation of a
rotating p-dimensional surface embedded in a sphere $S^{d-2}$ of time
dependent radius $r(\tau)$. 

In this case, the parton metric has the form
\bea
&&q_{ab}=r^2 \partial_a{\bf m}\cdot\partial_b{\bf m}\equiv 
r^2 {\tilde q}_{ab},\;\;\;q=r^{2p}{\tilde q},\;\;\;a,\,b=1,\cdots,p,\n\\
&&h_{\mu\nu}=[D{\tilde h}D^T]_{ij},\;\;\;
{\tilde h}_{ij}\equiv -{\tilde q}^{ab}\partial_a m_i \partial_b m_j,
\;\;\;i,\,j=1,\cdots,d-1,
\eea
where ${\tilde q}^{ab}$ is the inverse of ${\tilde q}_{ab}$ and 
${\tilde q}=\mbox{det}{\tilde q}_{ab}$.
Taking the rotation matrix as
\be
\l{rm}
D(\tau)=\exp[\varphi(\tau)A],
\ee
where the matrix $A$ is antisymmetric, the Hoppe-Nicolai solution
corresponds to the ansatz choosing ${\bf m}(\Omega)$ to be \c{hn}
\bea
\l{A2}
&& A^2 {\bf m}(\Omega)=-{\bf 1}\cdot {\bf m}(\Omega),\\
\l{ortho}
&&\partial_a {\bf m}^T A {\bf m}=0.
\eea
The above equations can be satisfied by choosing 
\be
\l{m}
{\bf m}=(n_1, n_2, \cdots, n_k, 0,\cdots,0),
\;\;\;\;p+1\le k \le \frac{d-1}{2},
\ee
and 
\be
\l{A}
A=\pmatrix{ 0 & -{\bf 1} & 0 \cr
                    {\bf 1}  & 0 & 0 \cr
                     0 & 0 & 0 \cr},
\ee
where ${\bf 1}$ is the $k \times k$ unit matrix. 
Then the p-brane action for the
pulsating and rotating surfaces also takes the simple form
\bea
\l{rsp}
I&=&\frac{A_p}{2}\int d\tau\, r^p[{\tilde e}^{-1}(-{\dot t}^2 
+{\dot r}^2+r^2{\dot \varphi}^2)-{\tilde e}]\n\\
&=& -A_p\int d\tau\, r^p \sqrt{{\dot t}^2 -{\dot r}^2-r^2{\dot \varphi}^2},
\eea
where $A_p=\int_\Sigma d^p\sigma \sqrt {\tilde q}$ is the area of
$p$-dimensional surface embedded in a $(k-1)$-dimensional unit
sphere. In deriving Eq. (\r{rsp}), the terms involved with $h_{\mu\nu}$
identically vanish due to the Eq. (\r{m2=1}), Eq. (\r{ortho}) 
and the orthonormality relation $D^T(\tau)D(\tau)={\bf 1}$.

Notice that the variation of $A_p$, together with the constraint 
${\bf n}^2=1$, leads to the requirement that ${\bf n}$ describes 
a minimal surface in $S^{k-1}$ \c{hn}:
\be
\l{ms}
\nabla^2 {\bf n}(\Omega)=-p\, {\bf n}(\Omega),
\ee
where $\nabla^2=(1/\sqrt{\tilde q})\partial_a \sqrt{\tilde q} 
{\tilde q}^{ab}\partial_b$.

The equations of motion obtained by the variations $\delta t$ and 
$\delta \varphi$, respectively, are given by
\bea
\l{eomp}
\ba{l}
\partial_\tau \left(\frac{r^p\, {\dot t}}
{\sqrt{{\dot t}^2 -{\dot r}^2-r^2{\dot \varphi}^2}}\right)=0,\\
\partial_\tau \left(\frac{r^{p+2}\, {\dot \varphi}}
{\sqrt{{\dot t}^2 -{\dot r}^2-r^2{\dot \varphi}^2}}\right)=0.
\ea
\eea
Taking into account of $\tau$-reparameterization symmetry 
in the action (\r{rsp}), the above equations of motion take the form
of the energy and the angular momentum conservation, respectively,
\bea
\l{eacp}
\ba{l}
{\dot r}^2+r^2{\dot \varphi}^2
+\alpha^2 r^{2p}=1,\\
r^2{\dot \varphi}=\mbox{constant}\equiv L,
\ea
\eea
where $\alpha=\sqrt{1-L^2/r_0^2}/r_0^p$ and 
$r_0$ is the radial position at $\tau=0$. 
From Eq. (\r{eacp}), it follows that
\be
\l{p=1}
{\dot r}^2+\frac{L^2}{r^2}+\alpha^2 r^{2p}=1,
\ee
which is compatible with the equation of motion determined by 
variation $\delta r$.
For the case of $L=0$ and $p=2$, the Eq. (\r{p=1}) equals to
the Eq. (\r{ec}) for the spherical membrane.
In the case of $L=0$, there is no need to put the restriction on 
${\bf m}(\Omega)$ such as Eqs. (\r{A2}) and (\r{ortho}).

The other solutions of Eq. (\r{p=1}) are obtained by straightforward
integration 
\be
\l{solp}
\tau=\frac{1}{2}\int \frac{dz}{\sqrt{z-\alpha^2 z^{p+1}-L^2}},
\ee
where $z=r^2$. For $p=1$, $r(\tau)=(1/\sqrt 2 \alpha)
\sqrt{1+2\alpha a \sin(2\alpha \tau +\theta_0)}$, where
$a=\sqrt{1/4\alpha^2 - L^2}$. For $p=2,\,3$, the solutions can be
also solved by elliptic functions \c{GR}. 
They describe the motion of partons pulsating by $r^{2p}$ potential 
with angular momentum $L$.
Note that, for $L\neq 0$ and
finite energy, the pulsating and rotating p-branes need not collapse
to a point, that is, there is a nonzero minimum radius $r_{min}$
determined by Eq. (\r{p=1}).

\section{Discussion}

The aim of this paper is to understand p-brane dynamics in terms of 
superparticles. Although the parton picture in terms of 
superparticles is quite different
from those of Matrix theory and string bits model, 
we have found that the super p-brane dynamics can be understood by the
collective dynamics of superparticles in a unified framework. 
Here we summarize our formulation and add some comments on the matrix 
formulation of the supermembrane. 

The Matrix formulation of supermembrane was constructed 
according to the following scheme. 
In light-cone gauge, the residual reparameterization symmetry 
reduces to an area preserving diffeomorphism, $SDiff(\Sigma)$. 
According to the relation between the representation of $sdiff(\Sigma)$
and the $N \rightarrow \infty$ limit of some Lie algebra \c{kr}, 
the light-cone supermembrane is mapped to a {\it physically
equivalent} system with the corresponding gauge symmetry. 
Here, {\it physically equivalent} means that the physical degrees of
freedom and their Hilbert space structure exactly match with each
other. Interestingly, such a system exists and is given by a
supersymmetric Yang-Mills quantum mechanics \c{whn}. When this is done, 
the embedding coordinates $X^\mu(\tau,\sigma^a)$ and 
$\theta(\tau, \sigma^a)$ are mapped to matrices $X^\mu_{IJ}(\tau)$ 
and $\theta_{IJ}(\tau)$ transforming in the adjoint representation 
of the Lie group $G$. The $\Sigma$-dependences of $X$ and $\theta$ 
are transformed to matrix degrees of freedom. 
That is, the matrix coordinates $X$ and $\theta$ are the
collective variables describing the many point-like parton 
degrees of freedom. The important point is that the matrix 
regularization of membrane dynamics is performed in a 
supersymmetric way. 

It is the recent picture of Matrix theory \c{BFSS,Matrix}
that some spectrums of M-theory in infinite momentum frame can be 
understood as the collective excitations of D0-particles whose
dynamics is given by a matrix quantum mechanics. 
In the BBS action of supermembrane in terms of superparticles, 
the $\Sigma$-dependences are collected into the form of the ``effective
potentials'', $G_{\mu\nu}$ and $\Pi_A$, between superparticles 
and summed over all constituent superparticles. 
The $Diff(\Sigma)$ symmetry restricts the form of the effective potentials. 
In other words, they should be given by the $Diff(\Sigma)$-invariants 
such as the MPB. Moreover, the $\kappa$-symmetry determines the
``gauge potentials'' $\Pi_A$ coupled to the superparticles. 
These potentials determine an effective background about
superparticle dynamics. Speculatively, the full matrix formulation of
supermembrane may reduce to a problem to encode the effective
background geometry determined by the potentials $G_{\mu\nu}$ and 
$\Pi_A$ into the collective (matrix) coordinates of superparticles. 

The possibility of a covariant (in the sense of the target space) 
matrix formulation rests on whether or not we can find a 
{\it physically equivalent} system with supersymmetric matrix
regularization that the dynamical degrees of
freedom and their Hilbert space structure exactly match with each
other. As pointed out by Smolin \c{smolin}, 
the only $SDiff(\Sigma)$ is linearly 
realized by the Poisson algebra (\r{mpoisson}), which is mapped to
the Lie algebra of a gauge group in light-cone gauge. The area
non-preserving part, $Diff(\Sigma)/SDiff(\Sigma)$, is non-linearly
realized by the Poisson algebra. If we want to have a covariant matrix
formulation of membrane, we should find a matrix realization 
(regularization) of the full $Diff(\Sigma)$ \c{smolin,am}. 
It is desirable in the practical sense that the matrix formulation
would provide the linear realizations on 
the $Diff(\Sigma)$, $\kappa$-symmetry and supersymmetry. 
Unfortunately, it seems that there is no definite 
recipe for the above issues at the moment. 

We think that, if the full matrix formulation of supermembrane should be
incorporated with all the recent pictures appeared in the
nonperturbative string theory and M-theory \c{Matrix,schtown}, e.g., 
noncommutative spacetime geometry, holographic principle, and p-brane
democracy, it will need a fundamental unit defining spacetime quanta,
bits of information, and partons of p-brane. 
We hope, in this sense, that the reformulation of p-brane dynamics 
by smaller entities
presented in this paper will be helpful to understanding 
the nonperturbative dynamics of the supermembrane. 

\section*{ACKNOWLEDGMENTS}
This work was supported by the Korea Science and
Engineering Foundation through Center for Theoretical Physics and 
by the Korean Ministry of Education (BSRI-98-2414).

\end{document}